\documentclass[runningheads]{llncs}
\usepackage{amsmath,amssymb,amsfonts}
\usepackage{dsfont} 
\usepackage{url} 
\usepackage{xcolor} 
\usepackage{cleveref}
\usepackage{cite}
\usepackage{mwe}
\usepackage{graphicx} 
\usepackage{tabularx}
\usepackage{multirow} 
\usepackage{booktabs}

\newcommand{\newtext}[1]{#1} 

\newcommand{\Domain}{\operatorname{\Omega}}
\newcommand{\DirSet}{\operatorname{K}}

\newcommand{\OriginalImage}{f} \newcommand{\dOriginalPixel}{f}
\newcommand{\dOriginalImage}{\boldsymbol{\dOriginalPixel}} 
\newcommand{\RecImage}{u}

\newcommand{\dRecPixel}{u} \newcommand{\dRecImage}{\boldsymbol{\dRecPixel}}
\newcommand{\DirFunc}{g} \newcommand{\dDirFunc}{\boldsymbol{g}}
 
\newcommand{\BasisMatrixElement}{B}

\newcommand{\BasisMatrix}{\boldsymbol{\BasisMatrixElement}}

\begin{document}
\title{Efficient Data Optimisation for Harmonic Inpainting with 
Finite Elements\thanks{This project has received funding from the European Research Council 
(ERC) under the European Union's Horizon 2020 research and innovation 
programme (grant agreement No 741215, ERC Advanced Grant INCOVID).}}
%
%
\author{Vassillen Chizhov \and Joachim Weickert} 
\authorrunning{V. Chizhov \and J. Weickert} 
%
\institute{Mathematical Image Analysis Group, 
Faculty of Mathematics and Computer Science,\\ 
Campus E 1.7, Saarland University, 66041 Saarbr{\"u}cken, Germany, \\
\email{\{chizhov, weickert\}@mia.uni-saarland.de}}

\maketitle              
\begin{abstract}
Harmonic inpainting with optimised data is very popular 
for inpainting-based image compression. We improve this 
approach in three important aspects.  Firstly, we replace the 
standard finite differences discretisation by a finite element method 
with triangle elements. This does not only speed up inpainting and data 
selection, but even improves the reconstruction quality. 
Secondly, we propose highly efficient algorithms for spatial and tonal 
data optimisation that are several orders of magnitude faster than 
state-of-the-art methods. Last but not least, we show that our 
algorithms also allow working with very large images. This has
previously been impractical due to the memory and runtime requirements 
of prior algorithms.

\end{abstract}
%
%
%

\section{Introduction}

In recent years, alternatives to transform-based
compression have been proposed under the name inpainting-based
compression; see e.g.~\cite{Galick2005, Demaret2006, Liu2007, Mainberger2011, 
Hoeltgen2013, SPME14, Chen2014, Karos2018, Daropoulos2020, Andris2021b}. 
During encoding, these approaches store an optimised subset of 
the image data (e.g. $5\%$ of all pixels), the so-called \textit{inpainting
mask}. In the decoding phase, they 
reconstruct an approximation of the original image from the mask data 
with the help of an inpainting process. 
Inpainting-based methods have been able to qualitatively outperform
widely used codecs such as JPEG and JPEG2000; see e.g.~\cite{SPME14}.
Two of the simplest inpainting methods are based on linear
spline approximation over triangle
meshes~\cite{DNV97,Demaret2006,Adams2013,Marwood2018} 
and on discretisations of the Laplace equation~\cite{Mainberger2012}
(also called {\em homogeneous diffusion inpainting} or 
{\em harmonic inpainting})
with the finite difference method (FDM). Although these approaches are 
relatively simple, they can achieve very good quality if the inpainting 
data are carefully optimised~\cite{Demaret2006, Mainberger2012, Hoeltgen2013, 
OCBP14, BLPP17}. Their quality also exceeds the one reported for recent
neural network approaches for sparse inpainting~\cite{Dai2020}.


Unfortunately, the data optimisation during the encoding phase is 
typically costly both in terms of memory and runtime. It consists of two
problems: spatial optimisation and tonal optimisation. The spatial optimisation
problem aims at finding the optimal locations on the image grid for the set
of pixels to be stored. This is a hard combinatorial 
minimisation problem that requires efficient heuristics.
In contrast, the tonal optimisation problem consists of modifying the 
grey values (or colour values) of the stored data. While this is a 
least squares problem,
it can require large computational and memory 
resources~\cite{Mainberger2012,Hoeltgen2015,Hoeltgen2017}. This
precludes previous tonal optimisation methods for harmonic inpainting
from being applied to very large images.

Artifacts constitute another problem. Linear spline inpainting may suffer
from conspicuous structures of the triangle mesh, and FDM-based harmonic 
inpainting typically exhibits logarithmic singularities at the stored 
pixels.


\paragraph{\normalfont \textbf{Our Contributions.}} 

The goal of our paper is to address the above-mentioned problems by
introducing three improvements of high practical relevance:

\begin{enumerate}

\item We propose to implement harmonic inpainting with a finite element
  method (FEM) based on adaptive triangulation. Compared to finite 
  differences on the pixel grid, its adaptivity allows to achieve better 
  quality in lower time.  Interestingly, this also alleviates the 
  logarithmic singularities that are characteristic for the finite 
  difference solution.

\item We devise a computationally efficient spatial optimisation strategy 
  which scales better and is multiple orders of magnitude faster than 
  current state-of-the-art approaches without compromising reconstruction 
  quality. 

\item We propose a computationally and memory efficient algorithm for tonal 
  optimisation. Its runtime and memory requirements scale much more 
  favourably with the image resolution compared to previous state-of-the-art 
  approaches. This allows us to handle also very large images, which 
  has been problematic for previous tonal optimisation approaches 
  for harmonic inpainting. 

\end{enumerate}


\paragraph{\normalfont \textbf{Related Work.}} 

Finite elements have been successfully used for PDE models for image 
denoising~\cite{KM95,BMi96,Preusser2000} and
restoration~\cite{Boujena2015, Theljani2017}. However, to our knowledge 
they have not been applied to PDE-based image approximation from sparse data. 

Various spatial optimisation strategies have been proposed for
inpainting-based compression.
Mainberger et al.~\cite{Mainberger2012} introduced a probabilistic 
sparsification approach that selects a subset 
of pixels to be removed at each iteration. A subsequent
nonlocal pixel exchange step relocates pixels probabilistically 
and keeps only relocations that decrease the error.
The dithering-based densification strategy of Karos et al.~\cite{Karos2018} 
iteratively adds a fraction of the target pixels by halftoning an 
inpainting error image. 
The Voronoi densification of Daropoulos et al.~\cite{Daropoulos2020} 
constructs a Voronoi diagram from the currently stored pixel data and 
inserts a new pixel in the cell with the highest error at each iteration. 
Our approach combines the ideas of densification with an error 
map~\cite{Karos2018} 
and a partitioning of the domain~\cite{Daropoulos2020}. However, we use 
the tessellation already available from the FEM mesh. In our case this 
is a Delaunay triangulation which is dual to a Voronoi tessellation and
avoids storing the mesh connectivity.

Tonal optimisation approaches can be split into two categories: 
methods that explicitly compute the dense solution for the inpainting, 
and iterative descent-based strategies. The former ones include 
inpainting echoes~\cite{Mainberger2011}, LSQR relying on an LU 
factorisation~\cite{Hoeltgen2015}, and Green's functions~\cite{HoffmannPhD}. 
Their memory requirements scale quadratically in the 
number of mask points or image pixels. This makes them inapplicable for 
larger images. 
The second category includes gradient descent methods~\cite{HoffmannPhD,
Hoeltgen2017}, quasi-Newton techniques such as L-BFGS~\cite{Chen2014},
and primal-dual algorithms~\cite{Hoeltgen2015}. Our approach falls in this
category and is very efficient by using nested conjugate gradient 
iterations~\cite{Sa03}.


\paragraph{\normalfont \textbf{Outline.}} 

In~\Cref{ss:FEM_review} we discuss the mathematical formulation of 
harmonic inpainting and the application of FEM to it. Then we describe 
our spatial optimisation approach in~\Cref{ss:spatial_optimisation} and our
tonal optimisation algorithm in~\Cref{ss:tonal_optimisation}. We present 
our results in~\Cref{ss:experiments}, and we conclude the paper with
 some potential avenues for future research in~\Cref{ss:conclusion}.


\section{FEM for Harmonic Inpainting} \label{ss:FEM_review}

In this section we introduce the mathematical formulation for
harmonic inpainting in the continuous setting and then 
motivate the choice of an FEM discretisation over an FDM discretisation.
Additionally, we describe how the FEM mesh is constructed and how the 
solution is interpolated to all image pixels.


\subsection{Continuous Formulation of Harmonic Inpainting}

We use harmonic inpainting to construct an approximation of the original 
image given a sparse subset of the image data. Let us consider
a continuous greyscale image $\OriginalImage : \Domain 
\rightarrow \mathbb{R}$ on some rectangular image domain $\Domain$. 
Rather than storing $f$ on the entire domain $\Domain$, we only specify
values on a {\em data domain} $\DirSet \subset \Domain$ (the so-called
{\em mask pixels} in a discrete setting). In the remaining domain
$\Domain \setminus \DirSet$, we fill in missing values
by solving the Laplace equation with reflecting boundary conditions
on $\partial\Domain$:
\begin{alignat}{3}
 -\operatorname{\Delta}\RecImage(\boldsymbol{x}) &= 0, \quad &\boldsymbol{x}& \in
   \Domain \setminus \DirSet, \\ 
 \RecImage(\boldsymbol{x}) &= \DirFunc(\boldsymbol{x}), \quad &\boldsymbol{x}&
   \in \DirSet, \\ 
 \partial_{\boldsymbol{n}} \RecImage(\boldsymbol{x}) &= 0, \quad &\boldsymbol{x}&
   \in \partial\Domain,
\end{alignat}
where $\boldsymbol{n}$ is normal to $\partial\Domain$. In order to improve 
the approximation quality of $u$ w.r.t. $f$, one optimises the shape 
of the data domain $\DirSet$ under some size constraint {\em (spatial 
optimisation)} and the corresponding grey values $g$ within $K$ {\em 
(tonal optimisation)}.
In the discrete setting, $\dOriginalImage$, $\dDirFunc$, and $\dRecImage$ 
are vectors instead of functions.


\subsection{FEM Formulation}

Finite difference methods (FDM) \cite{MG80} and finite element methods (FEM) 
\cite{Johnson2009} are two classes of numerical techniques used to solve 
differential equations.
FDM is often applied on a regular and equidistant grid (the pixel grid for
digital images), while FEM is well-suited for data on adaptive 
meshes. Therefore, the number of unknowns for the FDM method 
grows rapidly with the image resolution, while one has fine control over the 
number of unknowns in FEM by means of the construction of the underlying mesh.
This flexibility naturally suggests using an adaptive lower resolution FEM 
mesh with fewer unknowns in order to speed up the inpainting compared to 
FDM. 

Our FEM formulation relies on a triangle mesh; see Figure 
\ref{fig:Dirichlet_peaks}, right. 
Every mask pixel is a vertex in this mesh, and a subset 
of the remaining non-mask pixels are chosen as unknowns, depending on 
how many unknowns are required or on the runtime constraints. We call
this subset the \textit{unknown vertices}. 
If all non-mask pixels were chosen as unknown vertices, then the 
standard (5-point stencil) FDM discretisation of the Laplace equation would 
be recovered.
Once the vertices for the mesh have been determined, a Delaunay triangulation 
is constructed from the point set formed by the vertices~\cite{Preparata1985}.
The Delaunay property of maximising the minimum angle in the triangulation 
is desirable, since it provides guarantees regarding the optimality of the 
resulting condition number for the matrices involved in the inpainting and 
tonal optimisation~\cite{Shewchuk2002}.
Furthermore, this frees us from storing any connectivity, since the mesh 
can be reconstructed only from the mask pixels and unknown vertices~\cite{Preparata1985}.

We consider a {\em linear} FEM method: The reconstructed image is linear 
within each triangle of the aforementioned mesh and continuous at edges 
and vertices. At each mask pixel the corresponding stored grey value 
is prescribed, while the values at the unknown vertices are found by solving 
the linear system arising from the FEM formulation. Each unknown corresponds
to an unknown vertex. We typically choose as many unknown vertices as we 
have mask pixels. For further technical details on FEM, we refer to the 
standard literature~\cite{Johnson2009}. Given a Delaunay mesh with at 
least one mask pixel, the system matrix is symmetric and positive definite. 
Thus, we can use the conjugate gradient method~\cite{Sa03} in order to 
approximate the solution. Finally, the solution is linearly interpolated 
within each triangle for the remaining non-mask pixels which do not 
correspond to vertices in the mesh.


\section{Spatial Optimisation} \label{ss:spatial_optimisation}

Our spatial optimisation algorithm is based on a novel coarse-to-fine
error map densification strategy that combines concepts from an error 
map dithering~\cite{Karos2018} and a Voronoi densification
algorithm~\cite{Daropoulos2020}. The main idea is to introduce $m$ 
mask pixels over $n$ iterations, where each iteration requires a
single inpainting.

To this end, we have investigated many strategies, and we now describe 
the one which has produced the best results. Most importantly, we have 
found that the error map must be considered at a locally adaptive
scale that reflects the mesh structure as a function of the mask pixel 
density. In each iteration we 
introduce $\frac{m}{n}$ of all mask pixels. In the very first iteration 
we distribute all unknown vertices uniformly at random on the image 
grid, and introduce the first $\frac{m}{n}$ mask pixels also 
uniformly at random. 
In each subsequent iteration, we compute the inpainted image $\dRecImage$  
from the current mesh, and its error map 
$e_i = (\dRecPixel_i-\dOriginalPixel_i)^2$ in each pixel $i$. 
Afterwards we evaluate the total $L_2$ error in each triangle. We
insert a single mask pixel in each triangle, in descending order of 
the $L_2$ error. The position of the mask pixel within a triangle is 
chosen to be at the empty location with the largest pointwise error.

This procedure ensures that the inserted mask points are adapted to the 
evolving FEM mesh, which is the main advantage over previous approaches
~\cite{Karos2018,Daropoulos2020}.
Generally, the more iterations/inpaintings $n$ are allowed, the higher the 
quality of the mesh is for the purpose of approximating $\dOriginalImage$.
Practically useful numbers range between $n=10$ and $n=100$.


\section{Tonal Optimisation} \label{ss:tonal_optimisation}

Let the grey values in the $m$ optimised mask pixels be given by a 
vector $\dDirFunc \in \mathbb{R}^m$. Then the corresponding inpainting
result $\dRecImage$ can be written formally as 
$\dRecImage=\BasisMatrix \dDirFunc$, where the matrix $\BasisMatrix$
is dense and depends only on the mesh and the enumeration of the vertices.
It also includes the discussed interpolation 
from~\Cref{ss:spatial_optimisation} for image pixels that are not vertices 
in the mesh. The goal of the tonal optimisation problem is to find the
grey values $\dDirFunc^*$ that give the best reconstruction 
$\dRecImage^*=\BasisMatrix \dDirFunc^*$:
\begin{equation}
  \dDirFunc^* =\, 
  \underset{\dDirFunc\in\mathbb{R}^m}{\arg\min}\, \|\BasisMatrix \dDirFunc 
  - \dOriginalImage\|^2_2\,.
\end{equation} 

Since this is a linear least squares problem, its solution is given by  
the linear system of equations (the so-called normal equations): 
\begin{equation}
  \label{eq:tonal_optimisation_normal_equations} 
 \BasisMatrix^T \BasisMatrix \, \dDirFunc =
 \BasisMatrix^T \, \dOriginalImage.
\end{equation} 
A straightforward approach to solve the normal equations would compute
$\BasisMatrix$ explicitly. Unfortunately, this would require memory that 
scales {\em quadratically} in the number of pixels. This becomes 
unfeasible for very large images.

As a remedy, we exploit the fact that $\BasisMatrix$ is formed as the 
product of a sparse matrix (interpolation) and the inverse of a sparse 
matrix (Laplace equation). This means that the required memory for our 
optimisation scales {\em linearly} in the number of pixels. We achieve 
this by applying a nested conjugate gradient solver to 
(\ref{eq:tonal_optimisation_normal_equations}). The inner iterations are
effectively solving inpainting problems due to the Laplace equation instead
of inverting the matrix from the FEM formulation explicitly. The outer 
iterations optimise the grey value vector $\dDirFunc$.

Interestingly, our approach results not only in a better memory scaling:
Also its runtime is much lower than the inpainting echo 
approach~\cite{Mainberger2012}, and it is also faster than alternative tonal
optimisation techniques~\cite{Hoeltgen2017,HoffmannPhD} for usual
mask densities.


\begin{figure}[tb]
\begin{center}
\begin{tabular}{c c c}
 original image \textit{trui} &
 original image \textit{walter} &
 original image \textit{peppers}\\[1mm]
\includegraphics[width=0.3\textwidth]{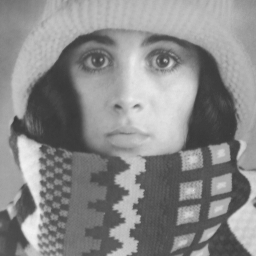} 
\hspace{0mm} & \hspace{0mm}
\includegraphics[width=0.3\textwidth]{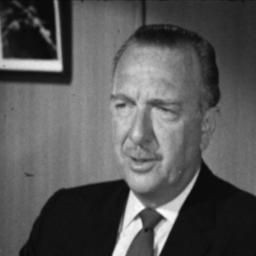}  
\hspace{0mm} & \hspace{0mm}
\includegraphics[width=0.3\textwidth]{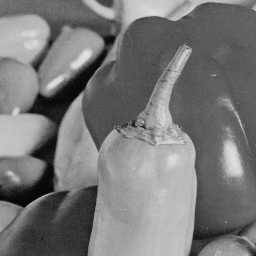}\\[2mm] 
\includegraphics[width=0.3\textwidth]{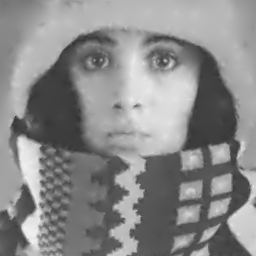} 
\hspace{0mm} & \hspace{0mm}
\includegraphics[width=0.3\textwidth]{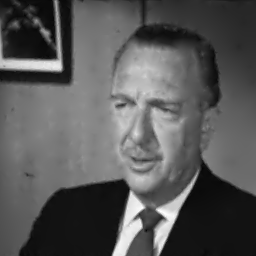} 
\hspace{0mm} & \hspace{0mm}
\includegraphics[width=0.3\textwidth]{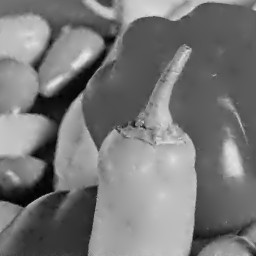}
\\[0.5mm]
FEM reconstruction & 
FEM reconstruction & 
FEM reconstruction 
\end{tabular}
\end{center}

\caption{FEM reconstructions for \textit{trui}, \textit{walter}, and 
\textit{peppers} with 4\% density and $n=100$.
\label{fig:walter_peppers}}
\end{figure}


\begin{figure}[tb]
\begin{center}
\begin{tabular}{c c c}
\includegraphics[width=0.3\textwidth] {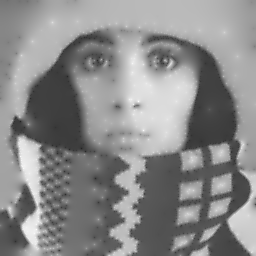} 
\hspace{0mm} & \hspace{0mm}
\includegraphics[width=0.3\textwidth] {images/trui_04_ours100_ton.png}  
\hspace{0mm} & \hspace{0mm}
\includegraphics[width=0.3\textwidth] {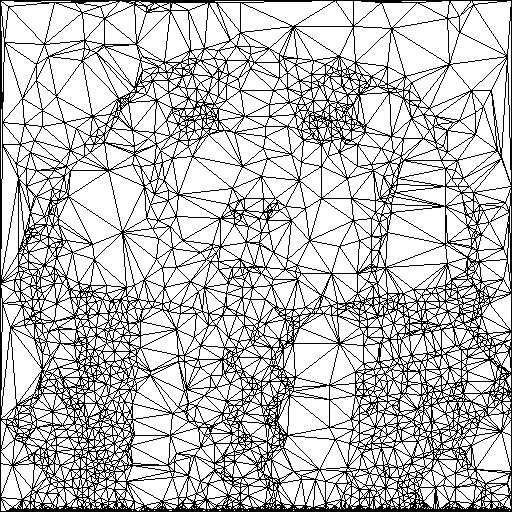}\\[0.5mm]
 FDM,\, MSE=36.04 & FEM, \, MSE=26.62  & \hspace{1mm} Delaunay triangulation
\end{tabular}
\end{center}

\caption{
Comparison between FDM with probabilistic sparsification (left) and FEM 
with densification (middle), both at 4\% density and tonally optimised. 
The FEM result does not suffer from the pronounced singularities on the 
scarf and face present for FDM.
Right: Delaunay triangulation used for the FEM method.
\label{fig:Dirichlet_peaks}}
\end{figure}


\begin{table}[b]
    \centering
    \caption{MSE comparisons at 4\% density without and with tonal optimisation (TO).}
    \label{tab:mse}
    \begin{tabularx}{\textwidth}{X c c | c c | c c}
        \toprule
        \multirow{2}{*}{\textbf{Method}} &
        \multicolumn{2}{c}{\textit{trui}} &
        \multicolumn{2}{c}{\textit{walter}} &
        \multicolumn{2}{c}{\textit{peppers}} 
        \\
        \cmidrule(lr){2-3}\cmidrule(lr){4-5}\cmidrule(lr){6-7} &
        \, no TO \, & with TO \, & 
        \, no TO \, & with TO \, & 
        \, no TO \, & with TO
        \\
        \cmidrule{1-7}
        Prob. spars. ($q=10^{-6}$) &
        66.11 & 36.04 &
        32.96  & 19.24 &
        44.85  & 28.58
        \\
        Ours ($n=10$) &
        44.62 & 30.07 &
        19.09 & 12.62 &
        43.20 & 29.83
        \\
        Ours ($n=30$) &
        40.58 & 28.21 &
        16.35 & {\bf 11.09} &
        38.37 & {\bf 28.11}
        \\
        Ours ($n=100$) &
        {\bf 37.60} & {\bf 26.62} &
        {\bf 15.92} & 11.21 &
        {\bf 36.68} & 28.85
        \\
        \bottomrule
    \end{tabularx}
\end{table}


\begin{table}[tb]
\centering


\caption{Runtime scaling with resolution (in seconds) of our
 spatial optimisation and tonal optimisation with $n=10$ at
 a density of 4\%.}
\label{tab:scaling}

\begin{tabular}{l c c c c c}
 \toprule
 \, image size \, & \,$64\times 64$ \, & \,  $128\times 128$ \, &   
 \,  $256\times 256$ \, & \, $512\times 512$ \, &   
 \, $1024\times 1024$ \, \\
 \cmidrule(lr){1-6}
 \, spatial optimisation \;\; & 0.01 & 0.03 & 0.11 & 0.49 & 2.07 \\
 \cmidrule(lr){1-6}
 \, tonal optimisation & 0.03 & 0.14 & 0.53 & 2.96 & 12.77 \\
 \bottomrule
\end{tabular}
\end{table}


\begin{figure}[tb]
\begin{center}
\begin{tabular}{c c}
\includegraphics[width=0.48\textwidth]
  {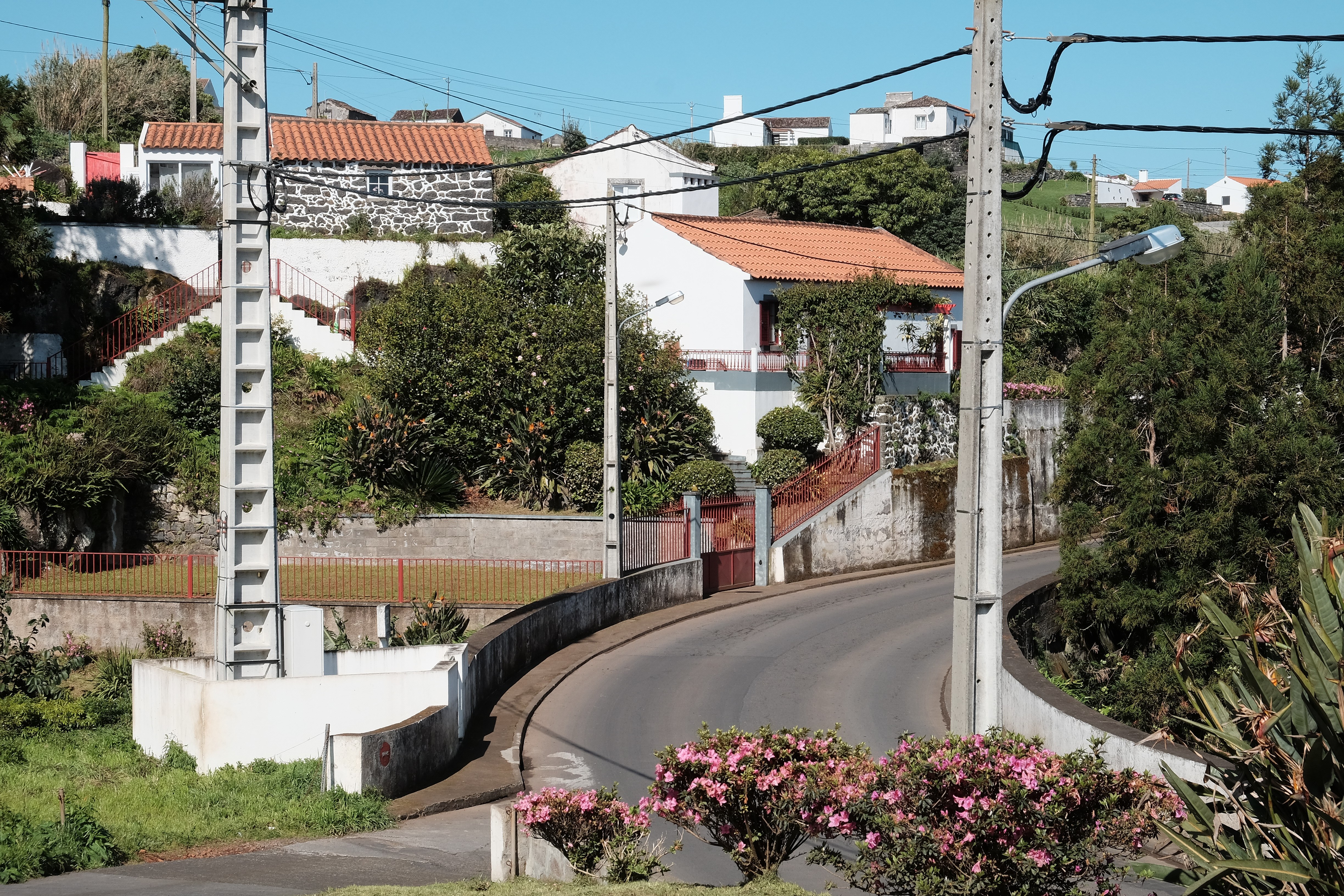} 
\hspace{0mm} & \hspace{0mm}
\includegraphics[width=0.48\textwidth]
  {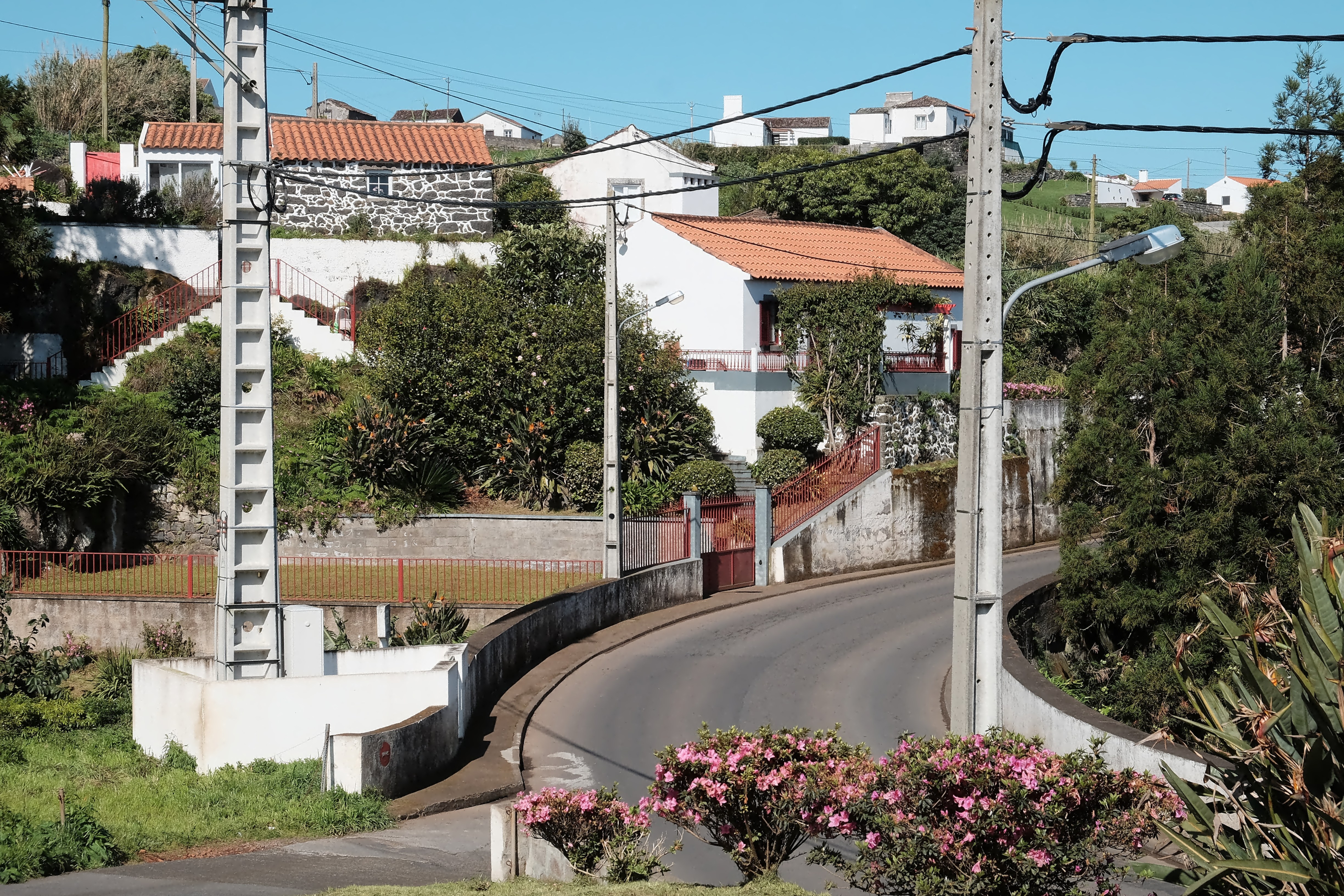}\\[0.5mm]
original image \textit{saomiguel} & \hspace{1mm}
 FEM reconstruction 
\end{tabular}
\end{center}

\caption{Left: A richly textured colour image of size $4896 \times 3264$ 
amounting to ca.~16 million pixels. Photo: J. Weickert.
Right: Our FEM reconstruction with $n=30$ iterations, 10\% mask 
density, and $L_1$ optimisation. 
\label{fig:image_16M}}
\end{figure}


\section{Experiments} \label{ss:experiments}


\Cref{fig:walter_peppers} displays the results of our FEM approach for
three classical test images that are popular for sparse harmonic 
inpainting: {\em trui}, {\em walter}, and (a $256 \times 256$ crop 
of) {\em peppers}.
We see that even at a low mask density of 4\%, the reconstructions
are fairly convincing. A quantitative analysis in terms of the mean squared 
error (MSE) is presented in~\Cref{tab:mse}, where we also compare our 
densification methods against the probabilistic sparsification results of 
Hoeltgen et al.~\cite{Hoeltgen2017}. We observe that our algorithms achieve a 
consistently better quality. When we apply tonal optimisation after the
spatial optimisation, the MSE decreases further by about one third.
Sometimes tonal optimisation can even fully compensate suboptimal
spatial optimisation, as can be seen for the {\em peppers} image.

In terms of runtime for the spatial optimisation, a probabilistic 
sparsification of a $256\times256$ image with the parameters 
from~\cite{Hoeltgen2017} on a Ryzen 4800H CPU takes about 10 minutes, 
while our FEM densification needs only 0.3 seconds. This shows that our 
method is 1800 times faster. This factor grows rapidly with the image size: 
For $512 \times 512$ images, it is already 10,000 (4 hours
versus 1.3 seconds). \newtext{The complexity of the optimisation is 
$\mathcal{O}(nq\sqrt{k})$, where $n$ is the number of iterations, 
$k$ is the condition number of the matrix, 
$q$ is the number of non-zero entries in the matrix 
(at most $6\times$ the number of vertices of the FEM mesh).}

Our tonal optimisation algorithm is also several orders of magnitude faster,
requiring 0.3s for a $256 \times 256$ image. The reported runtimes from 
Hoeltgen et al.~\cite{Hoeltgen2017} on a Xeon 3.2 GHz CPU vary between 
77s and 458s for the different algorithms discussed there. \newtext{The tonal 
optimisation has a time complexity of $\mathcal{O}(q\sqrt{k_2}\sqrt{k})$, where 
$k_2$ is the condition number of the tonal optimisation matrix.} Additionally, 
previous methods relying on QR or LU factorisations~\cite{Hoeltgen2015} 
require memory that grows quadratically in the number of mask pixels. 
In contrast, our algorithm's memory requirements grow linearly in the 
number of image pixels \newtext{($12q+32v$ bytes, where $v$ is the number of 
vertices in the FEM mesh)}. Thus, it can be applied to much larger images.

In \Cref{fig:Dirichlet_peaks}, we juxtapose FEM method with densification
to the FDM approach with probabilistic sparsification, 
both in the tonally optimised case. We observe that the FEM technique 
does not suffer from the pronounced singularities around mask pixels. 
These artifacts can be explained by the logarithmic singularities of 
the Green's functions in the continuous harmonic inpainting model; 
see \cite{MM12} for more details. Our FEM approach with linear 
approximations within each triangle handles them more gracefully.

\Cref{tab:scaling} reports scaling results for the spatial and
tonal optimisation steps of our FEM method. We see that it scales
almost linearly with the number of image pixels. The slight deviations
from an ideal linear scaling are caused by the conjugate gradient solver:
Large images lead to a higher condition number of the linear systems,
such that the conjugate gradient method needs a few more iterations 
\cite{Sa03}. 

Finally, \Cref{fig:image_16M} illustrates that our approach can also
reconstruct large and richly textured colour images with very high
perceptual quality. Extending the method to colour images is 
straightforward. In order to achieve highest visual fidelity, we have
followed \cite{Sinha2011} and optimised the $L_1$ error instead of
the $L_2$ one.

%
%

\section{Conclusions and Future Work} \label{ss:conclusion}

Our paper is the first one that considers finite elements for 
inpainting-based image compression. We have seen that they offer 
a much higher efficiency due to their better adaptivity to the 
image structure. Moreover, our algorithms have substantially
better scaling properties than previous approaches. This allows
to apply them to much larger images, which is an important issue
for bringing inpainting-based compression into practical applications.

In our ongoing work, we are studying options to increase the
efficiency of our approach even further and to extend it to
other PDEs and the compression of data on manifolds.



\bibliographystyle{splncs04}    
\bibliography{FEM_harmonic_inpainting}

\end{document}